# MINERVA: A PORTABLE MACHINE LEARNING MICROSERVICE FRAMEWORK FOR TRADITIONAL ENTERPRISE SaaS APPLICATIONS

Venkata Duvvuri[1]

[1]Oracle Corp &
[1]Department of Technology Leadership and Innovation, Purdue University, IL, USA
vduvvur@purdue.edu

## ABSTRACT

*In traditional SaaS enterprise applications, microservices are an essential ingredient to deploy machine learning (ML) models successfully. In general, microservices result in efficiencies in software service design, development, and delivery. As they become ubiquitous in the redesign of monolithic software, with the addition of machine learning, the traditional applications are also becoming increasingly intelligent. Here, we propose a portable ML microservice framework Minerva (microservices container for applied ML) as an efficient way to modularize and deploy intelligent microservices in traditional "legacy" SaaS applications suite, especially in the enterprise domain. We identify and discuss the needs, challenges and architecture to incorporate ML microservices in such applications. Minerva's design for optimal integration with legacy applications using microservices architecture leveraging lightweight infrastructure accelerates deploying ML models in such applications.*

## 1. INTRODUCTION

Enterprise SaaS applications are typically delivered as a service [1] to the client who need not worry about network, servers, operating systems, storage and data security. SaaS applications are broadly classified as general use and enterprise. The former involves general use software, such as Google Apps, and the latter specific enterprise applications, such as Oracle CX. Microservices are an effective way to decompose building large and complex systems into smaller sub-systems with these sub-systems interoperating via light weight (e.g., REST - representational state transfer) protocols. Machine learning sub-systems are increasingly important for SaaS applications like Oracle CX, Oracle CRM etc. due to the need to integrate intelligent decision-making. Many SaaS applications were built a decade or two ago, on an older technology stack on top of legacy data centers. Typically, this stack involves running monolithic applications on a multi-tenant SaaS infrastructure [2] with a huge database (like Oracle RDBMS) at its core. Pooyan et al. [3] identified several general benefits of microservices like faster delivery, improved scalability, and greater autonomy, turning an idea on some product manager's or other project member's whiteboard into a feature running in production as quickly as possible. Typically, microservices are packaged and deployed in the cloud using lightweight container technologies [4], following industry proven DevOps practices [5] and supported by automated software delivery machinery.

The paper is organized as follows: Section 2 introduces the machine learning needs in enterprise SaaS applications, Section 3 discusses related work, Section 4 elaborates how Minerva addresses these challenges, Section 5 highlights the system architecture, Section 6 & 7 compare and differentiate Minerva platform with other platforms, Section 8 presents the trade-offs made while designing Minerva, Section 9 & 10 presents Minerva's implementation and rollout in

Oracle and finally Section 11 & 12 concludes the paper with future research directions. Table 1 reflects the nomenclature used in this paper.

Table 1. Nomenclature

| Term | Description |
| --- | --- |
| CRM | Customer Relationship Management |
| CX | Customer Experience |
| REST | Representational State Transfer Protocol |
| SaaS | Software as a Service |
| RDBMS | Relational database Management System |
| DevOps | Development Operations |
| CPU | Central Processing Unit |
| RAM | Random Access Memory |
| ML | Machine Learning |
| B2B | Business to Business |

## 2. SAAS APPLICATION MACHINE LEARNING NEEDS

Machine learning models are designed by data scientists on sample datasets extracted for testing and experimentation. Typically, model development involves a fair understanding of the underlying optimization and statistical algorithms. A variety of programming languages, like R, Python etc., can be used along with various open source libraries for implementation. The typical tech-stack for development of these models is different from the host SaaS application stack.

The following additional requirements are not obvious, but they are crucial to implement machine learning in SaaS application. We highlight the technical requirements arising out of them. Section 6 and 7 showcases how Minerva achieves these technical requirements and as well suits the business needs, while others fall short.

### 2.1. Technical Requirements

- The need for *reusability* of the ML sub-system to serve numerous and diverse models to the host apps.

- The need for *decentralized* data governance and pre-processing to help with feature engineering required for machine learning models.

- The need for *scalability*, both horizontally (more machines) and vertically (add. CPU, RAM etc.).

- The need for real time or near real time (*online*) performance in predictions to serve intelligence in host applications.

- The need to accommodate long batch/*offline* training.

- The need to *secure* the sensitive data exchange between the feature processing subsystem and the machine learning processing sub-system.

- The need to build the ML microservice with the best of the breed modelling *polyglot* libraries in and *independent* of tech-stack of legacy system.

### 2.2. Business Requirements

Executives from SaaS product companies have established an aggressive 3 to 5 years horizon for cloud migration plans [6] for their customers. Some have an even longer sunset plan to stay competitive. Thus, companies like Oracle (and others) need an interim solution that fulfils the intelligence needs of traditional SaaS applications, especially in the B2B enterprise domain. Additionally, we capture the following business requirements that necessitates an innovative solution that is different from typical cloud machine learning solutions [7]:

- The need to serve intelligence in numerous traditional SaaS applications (e.g. SAP, Oracle etc.) in a *lightweight* fashion without much impact to the underlying systems or infrastructure. This market is roughly $20B revenue annually.

- The need to have ML "*algorithm/model run near the data*" as opposed to "move the data to algorithm/model". This arises due to the legal rules of various states/countries making it difficult for applications to migrate data outside their datacenters.

- The need to have an "interim" machine learning solution that is *compatible* with "legacy" SaaS applications, at-least for some time to come due to delays and inertia in adopting recent computing cloud techniques [7].

In a traditional enterprise SaaS Application and/or suite the following requirements are of premium importance: independence of the microservices programming stack, reusability, lightweight ML platform resource needs, algorithm running in proximity to data and compatibility with existing architecture without much impact.

## 3. RELATED WORK

Machine learning microservices have evolved from leveraging virtualization environments [8] to containerized approaches [9]. Ignacio et al. [8] has adopted a Bring Your Own Learner in the FCUBE project where various machine learning predictive algorithms can be run in a virtual machine environment [10]. They employ a plug and play approach which is a basic tenet of our approach as well. However, they focus on ensemble models for predictive classifications problems. Additionally, this approach is an offline approach and cannot be employed by traditional SaaS applications. Our solution generalizes to any kind of machine learning model. Fundamentally, our Bring Your Own Model or Algorithm (BYOMOA) approach allows for immense flexibility in choice of model libraries employed. Secondly, we also treat the model building exercise as a black box where the modeler obeys an established abstract interface of predict and train with our machine learning microservice framework. Pasquale et al. [9] has extended the FCUBE approach to a cCUBE microservices framework with containerization of services and adding orchestrators to help manage the compute units. While, we adopt such containerizing and task management as well, we offload management of the machine learning and data jobs to either out-of-box orchestrators or to the host application that allow such orchestrations. Unlike cCUBE we relax the limitations by designing for any supervised, unsupervised or deep learning algorithms. Thus, we allow almost unlimited ML capabilities in traditional SaaS applications.

Recently industry has recognized these issues by developing various ML platforms [11], [12], [14], [15] for ML lifecycle management. We borrow and extend their capabilities, and, in our case, it became essential due to value we need to deliver to SaaS applications suite. Similar to Databrick's MLFlow [11], we allow polyglot library capabilities by developing generic REST APIs that can use any ML library or algorithm. However, we offer a different deployment/microservices model by parcelling both model (M) and algorithm (A) in simpler project modules/files within container image instead of packaging models in complicated deployment repositories as in [11]. This allows for portability and enables wider "lift and shift strategy" that is the core value we deliver to SaaS application suites. Other recent ML platforms like Facebook's FBLearner [12], Uber's Michelangelo [13] and Google's TFX [14] have tried to solve the problem but within their own ecosystem. We concur with Zaharia et al. [11] that this limits ML developers to specific algorithms and libraries, decreasing their ability to experiment, and limiting ML developers and not allowing them to use new libraries or models.

Finally, the models have been traditionally integrated into products using hardcoded or embedded stacks. In the hardcoded approach the models are trained, and the model is captured as a mathematical or statistical function capable of being written directly into the host application during prediction process. In the embedded approach, the model is recoded into host application programming language. These two approaches suffer from the following deficiencies:

- Slower development process due to sequential or near waterfall development.
- Additional translation needed into the host application tech stack and/or programming language.
- Limited number of modelling libraries available in the host programming stack.
- Errors cropping up in the model translation efforts.
- No reuse of the model or code by related products in the traditional SaaS application.

## 4. ADDRESSING THE CHALLENGE

In our approach we devise a machine learning (ML) microservice sub-system "framework" (Minerva) within the ecosystem of a suite of traditional "legacy" SaaS applications. To address the needs of connected intelligence in such applications, we establish well defined REST service contracts with the various sub-systems. Due to the varied nature of interactions amongst the sub-systems we evolve the REST contracts into a "consumer contracts REST pattern" as suggested by Ian [15]. The machine learning model and/or its code is a black box that can be plugged into the Minerva by adopting well defined predict and train abstractions provided by the framework. To support monitoring and operations we allow for tiered logging interfaces that get ingested into the overall host application ecosystem by mounting a shared file system. Thus, we can adopt the same support operations to monitor the machine learning intelligence as the rest of the legacy ecosystem. A continuous deployment (CD) framework facilitates the agile development of Minerva parallel to the host application development. This eliminates a sequential development process and accelerates putting features into production. The cadence with data pre-processing jobs is orchestrated outside Minerva. These data jobs themselves can leverage database libraries already developed. Additionally, the very nature of independence of the microservices allows both the model and framework development in different programming languages and libraries than the "legacy" application ones. This adds to the flexibility in Minerva in choosing custom or advanced libraries for modelling. Organizationally, the microservice architecture allows a separate team to be responsible for this ML intelligence. This "engineering less" approach stems from the very conceptual adoption of ML microservices into the legacy development and deployment process. Another important contribution of our work is to allow for reuse and/or portability of the same machine learning model and/or algorithm code (A) by other related products (RP) in the parent suite organization. These products can now contribute ("lift and shift strategy") to mutually beneficial ML algorithms or models, enabled by Minerva's standard interfaces. These interfaces are served by a container that holds the model, its code (algorithm) and framework. The ease of adoption of Minerva in each RP is achieved by a separation of RP and framework configurations during deployment. An evolution of a given model is possible via a schema versioning in service payload contracts as evinced by Ian [15]. Another important contribution of the framework is the ability by the host applications to train and predict both online as well as offline. Finally, this containerized approach allows both vertical (CPU, RAM etc. of a node) and horizontal (spawning multiple nodes) scaling, albeit manually.

## 5. SYSTEM ARCHITECTURE

The system architecture and integrations are shown in Figure 1. The legacy application is built as several sub-systems: UI, DB, CORE, PLATFORM etc. in a datacenter. The ML microservices is a docker container (*Minerva*) implemented as per the proposal highlighted in section 4. Minerva container has three layers: *core*, *abstraction* and *app*. The Minerva core layer handles interaction with other systems, process management within the container, concurrency controls, security mechanisms and configurations. The Minerva abstraction stub wraps the algorithm/model, handles dynamic loading of projects (apps), versioning, exceptions and call-backs. The Minerva App layer implements the abstractions and codes the model, algorithm and logic using the best of breed ML libraries.

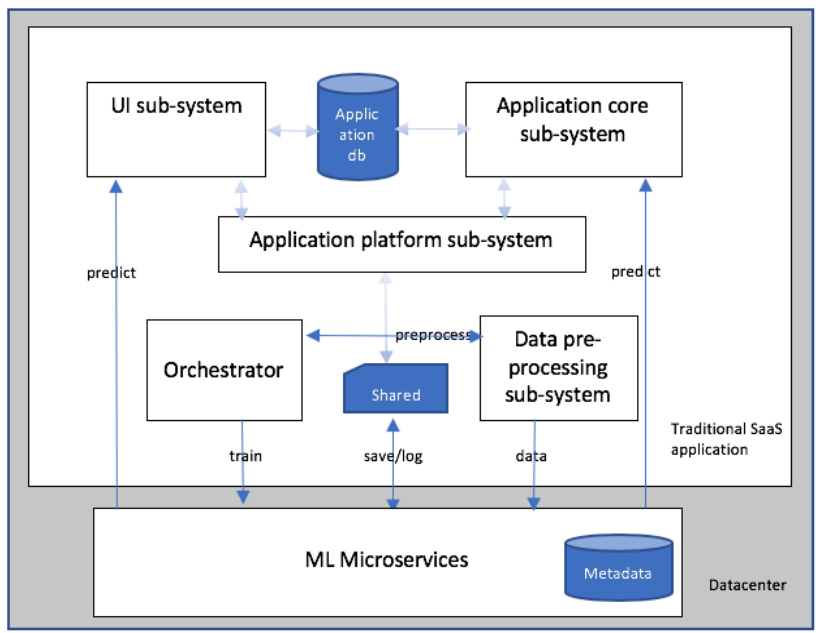

Figure 1. System architecture of ML enabled SaaS application

Minerva interacts with the legacy sub-systems like ui, core etc. for predictions. The training request can be orchestrated when the data is made ready for machine learning by the data processing unit. An orchestrator can be part of the legacy application, but that can also be pulled outside the application if needed. The data needed for ML is extracted and pre-processed by the data processing unit. The legacy application has a shared storage space that the ML microservice can leverage for saving and logging. The ML microservice is self-contained with its own meta-data info.

## 6. PLATFORM COMPARISON

Minerva platform is lightweight and primarily focuses on model/algorithm *lift and shift* enablement using a diverse set of ML libraries (TensorFlow, Pytorch, Keras, Sklearn etc.). In contrast, Kubeflow can serve fewer ML libraries (notably TensorFlow and Pytorch) and thus is not completely polyglot with respect to the choice of ML open source libraries. MLflow though polyglot and similar, has other design limitations.

Table 2. Comparison of Minerva with various platforms

| Feature | Minerva | MLFlow | Kubeflow |
|---|---|---|---|
| Job Tracking | X | X | X |
| Resources/ML Monitoring | | | X |
| Standard docker packaging | X | X | |
| Dynamic endpoints | X | | |
| API Standardization | X | | |
| Easy lift and shift of Algorithm | X | X | |
| Easy lift and shift of model | X | X | X |
| Microservices architecture (FaaS - Function as a Service) | X | X | |
| Real Time Serving (REST) | X | X | X |
| Batch processing (train) | X | X | X |
| Library Polyglot | X | X | |
| ML Pipelines | | | X |
| Model Visualization | | | X |
| Experimentation support | | X | X |
| Open Framework | X | | X |
| Model versioning | X | | |
| Concurrency control | X | | |

Secondly, Minerva focuses on ease of endpoint configurability which helps customizing the multitude of consumer driven REST APIs to interface with the various legacy application subsystems. These endpoint APIs are themselves standardized (for training) or templatized (for predictions) with version as a key attribute. Presumably, due to polyglot capabilities, these different models can be built using latest and greatest ML libraries in evolving versions, making avail of latest research. Thus, Minerva helps invoking multiple model revisions and swapping them as necessary by legacy application. MLflow has strengths in ML lifecycle management but offers little to help integrate with SaaS applications, especially with respect to API configurability, versioning or standardization. Although MLFlow has reusable projects, Minerva achieves algorithm *lift and shift* with simple modifications to pluggable single (or few) python project (*app*) files(s)/modules which are invoked dynamically by the framework. Thus, Minerva's approach of algorithm reuse and customization is different from MLFlow's heavy duty approach of external algorithm development involving separate and newer repositories conforming to complex templates. Kubeflow has a nicer set of complementary off-the-shelf capabilities like pipelines, scalability etc. which when supplemented to Minerva platform will enhance the overall robustness and maintenance of ML artifacts in SaaS applications. This integration with Kubeflow is illustrated in section 11 and earmarked for future studies of Minerva in OCI (Oracle Cloud Infrastructure) [16] platform. Finally, while scalability is handled in Minerva by using more and bigger containers, albeit replicating them manually, MLFlow's techniques of Spark [17] based compute and scalability does not bode well in legacy applications without adding elaborate infrastructure. We find that scalability requirements in legacy applications is not a severe requirement, considering the typical datasets in enterprise software.

# 7. VALUE DELIVERED TO LEGACY APPLICATIONS

Table 3 shows the pros and cons of Minerva adoption in legacy applications when compared to MLFLow and Kubeflow. Minerva optimally integrates with legacy subsystems, allows for democratization of algorithm development in a suite of products, leverages light weight infrastructure and is compatible with legacy architectures. We discuss these tenets below with comparisons.

## 7.1. Democratization

Minerva is designed for optimal portability of algorithms in a suite of applications leveraging the microservices architecture, standardizing APIs and enabling *lift and shift* mechanisms. While MLflow comes close, it is primarily designed for ML lifecycle management. It doesn't address techniques to simplify *lift and shift* needs of algorithms/models across applications in a suite. For this, MLFlow relies on building and importing a new code repository, while Minerva achieves it by easily modifying the algorithm pluggable project (*app*) file(s) in a docker image. Thus, the ingestion of a new algorithm code is lightweight and also seamless due to the dynamic loading of project modules by framework, thereby minimizing the time to customize algorithms. Thus, Minerva allows for democratization of model/algorithm development by several teams in the organization. However, MLFLow due to lack of configurability capabilities, makes this difficult. Secondly, one can directly import pretrained models into the new application in a suite, and thus need not train a new model per se for the new application in the suite. In all, Minerva is designed for easier *lift and shift* strategy which is a key enabler for democratization of algorithm development in an application suite.

Table 3. Pros and Cons of Minerva platform in legacy application suite

| Feature | Minerva | MLFlow | Kubeflow |
|---|---|---|---|
| Decentralized Microservices Architecture | X | X | X |
| Easier Integrations | X | | |
| Lightweight Infrastructure | X | | |
| Scalable Training | | X | X |
| Legacy Compatibility | X | X | |
| ML Democratization | X | | |

## 7.2. Integration

Integration is another differentiator in Minerva. With its standardized and self-documenting swagger end points [19] it can add, modify or delete them at will. Thus, design of multitude of consumer driven payloads [11] with various legacy application subsystems becomes standard and easy. Additionally, Minerva parses and validates payloads dynamically and supports API versioning and identifiers that can be used to respond back to calling subsystems. MLflow doesn't offer such configurability and hence is not easier to adopt or adapt when integrating with the many legacy application subsystems.

## 7.3. Compatibility

Kubeflow [18] can serve some needs of SaaS applications when they are migrated to cloud infrastructure due to its deep roots in kubernetes, a virtual cloud operation system. But, Kubeflow cannot be readily accessed or ingested by legacy applications as they will need an upgrade to the newer cloud infrastructure. Minerva not only works within the existing infrastructure with minimal computing requirements, but also does not need to move the data out of the legacy application's datacenter. It is compatible with legacy architecture and can easily integrate into them due microservices architecture.

Finally, the industry has variety of cloud ML platforms like Amazon AWS [20], Microsoft Azure [21] and Google Cloud [22]. They have ML Microservices enabled via REST end points, but the legacy applications cannot readily leverage them due to unwillingness to export data out and transfer into the respective cloud ecosystems. Additionally, the legacy applications may not migrate to these cloud platforms for some time to come as pointed in Section 2.

In all, configurability, containerization, lightweight infrastructure and standardization make Minerva quite portable as compared to MLFlow and Kubeflow.

## 8. DESIGN TRADE-OFFS

Minerva doesn't offer elastic scalability as yet, as we have traded off due to the inherent smaller sizes of datasets in enterprise SaaS applications, especially in enterprise domain. In contrast, Minerva relies on lighter infrastructure, and in order to alleviate scalability issues, it has concurrency controls via throttling mechanisms for every ML project.

Minerva has little to no support for ML pipelines. It has a rudimentary orchestrator leveraging some built-in support in Oracle CX products. This simpler approach was enough to support the existing workflows that have been built so far. However, Kubeflow [18] has better support for ML pipelines which Minerva could leverage as in Section 11 as and when the supported workflows become increasingly complex.

Minerva has primarily been designed to address deficiencies and delays in deploying ML. Hence, it leverages experimentation done outside the platform. With the addition of Kubefow's experimentation platform, Minerva could become more complete.

Finally, model explanation, visualization and performance monitoring are a needed features that Minerva lacks. This trade off was made due to build vs buy decisioning. Observing Kubeflow's capabilities there was no need to build these. Section 11 highlights that this along with other features can be brought into the Minerva with a mutually symbiotic integrations with one or more such platforms.

## 9. CASE STUDY

Minerva was successfully adopted for at-least four ML projects (apps) to build connected intelligence in Oracle CX product suite. The first ML project (app) was successfully implemented in one Oracle calendar release (three month's timeframe), the next three ML projects (apps) were adopted and integrated in the next release (subsequent three months). Thus, deploying ML intelligence features in production accelerated three times compared to traditional approaches with similar resources.

Consumer driven payloads form the foundation of the interaction with the sub-systems. We illustrate the case study with a few important ones. Figure 2 illustrates the design and implementation of a generic train payload for batch training any model. The above example specifically trains Customer Lifetime Value (CLV) model within CX product suite. Various job tracking info is passed in top section of the json payload. The framework captures job statuses, which help in monitoring and debugging issues. Also, a polling API exposes the statuses of the ML jobs to the orchestrator. The training happens at a pre-determined frequency per account (a.k.a customer) triggered by the host application using an orchestrator. The data is made available by the data processing unit at CLV_MODEL_INPUT table. The CLV algorithm will build and store the ML model in ML_MODEL_STORED table in a database. The database ensured the model security, versioning and fault tolerance. An asynchronous predict can be triggered similarly by changing action type to predict. This generates the results in

CLV_MODEL_OUTPUT for each account. The UI then displays the results to the end customers.

While, above use case has batch training and batch predictions, Minerva is not limited to these cases only. It can do online predictions, as well as hybrid predictions, where acknowledgements are synchronous and final predictions are batched (asynchronous) and reported in a call-back.

Figure 3 illustrates an online prediction of a new subject line (text) for consumption by the Ad/Message designer in CX product. This new endpoint can be easily created by swagger configs and additionally processing of its data attribute is handled by the downstream project (app) code/algorithm. Framework parses the remaining attributes to handle the versioning of APIs and synchronous replies within a required SLA (Service level agreement).

```json
{
    "accountId": 1234,
    "workflowExecId": 1,
    "workflowTaskId": "2",
    "workflowName": "CLV",
    "modelName": "model1",
    "actionType": "train",
    "modelInputPayload": {
        "tableNames": ["CLV_MODEL_INPUT"],
        "dbNames": ["DB1"]
    },
    "modelStoredTableName": {
        "tableNames": ["ML_MODEL_STORED"],
        "dbNames": ["DB1"]
    },
    "modelPredictedTableName": {
        "tableNames": ["CLV_MODEL_OUTPUT"],
        "dbNames": ["DB1"]
    },
    "serverName": "server",
    "dataSource": [{
        "dbUser": "db_service2",
        "dbURL": "user2",
        "dbPassword": "pwd2",
        "dbName": "pwd2"
    }],
    "sysAdminDbUrl": "db_service1",
    "sysAdminDbUser": "user1",
    "sysAdminEncryptedPwd": "pwd1"
}
```

Figure 2. Standardized payload to train ML model in Minerva

```json
{
    "version": "1",
    "accountId": "2",
    "data": {
        "subjectline": "50% off for new socks!"
    },
    "modelStorage": {
        "model": "abc"
    }
}
```

Figure 3. Online payload to predict subject line in Minerva

## 10. ORGANIZATION PERSPECTIVE

Organizationally, Minerva architecture allows for a separate data science(s) team to be responsible for crafting the algorithm, platform and the model. Not only one team, but several ML teams can contribute to the modelling activity, thus democratizing machine learning development. The infrastructure needed to run Minerva can be owned by separate engineering teams. This allows for data scientists to be responsible primarily for modelling and building ML pipelines. The data scientists can monitor their own pipelines and evaluate their own models. Additionally, the data science team can build a blueprint that serves as a deployment framework for engineering to deploy Minerva in many applications in a suite. This clever separation of responsibilities between the engineering and data scientists leverages their respective strengths. In many cases, international deployment teams can be made responsible for rolling out the blueprint to production further adding cost benefits.

## 11. FUTURE WORK

SOA (service-oriented architecture) [26] have continuously evolved since the initial days, with their adoption growing in cloud eco-systems. Minerva can work seamlessly in a new cloud infrastructure when applications migrate, especially Oracle cloud infrastructure (OCI) [23]. Notably, one can enable several other features along with Minerva platform in OCI. Specifically, OCI datascience module has capabilities to scale Minerva using Oracle Machine learning (OML) [24] which leverages compute power of the Oracle Autonomous database [25]. Additionally, more extensions to Minerva are possible due to native capabilities in OCI like Kubernetes [25] that can help build capabilities like dashboard monitoring, orchestration and account level debugging. Moreover, Minerva can draw strengths from OCI's Kubernetes pipeline engine - Kubeflow [19], its elastic machine scaling, its GPU compute power and its native load balancing capabilities. When the legacy applications migrate to cloud infrastructure (OCI or similar), one can leverage Minerva in an even better form. First version of Minerva rides out the interim period when legacy applications can't migrate to the cloud infrastructure.

## 12. CONCLUSION

Although microservices have conceptually existed since the days of SOA (service-oriented architecture), they have recently been adopted by various product organizations to reorganize monolithic SaaS applications. ML intelligence is a recent initiative to make these applications smart. Traditionally, the models built by data scientists have been integrated into products either using offline or embedded methods. We suggest a portable BYOMOA ML framework to allow for modeling flexibility, reusable ML use cases, agile development, tech-stack independence & faster deployments. This configurable and reusable *Minerva* framework is suited for legacy applications which can neither immediately migrate to cloud infrastructures nor can send their data outside their legacy datacenters for consumption by recent cloud ML platforms.

### ACKNOWLEDGEMENTS

Thanks to my wife and daughter for their encouragement.

Thanks to Oracle CX Marketing data science team members for their support.

Thanks to Dr. Michael Dyrenfurth - Purdue University, Dr. Nisha Talagala – Pyxeda, Dr. Sanjay Saigal - UC Davis and Dr. Kameshwar Yadavalli - Ostendo for their reviews.

# REFERENCES

[1] M. Ribas, A. S. Lima, N. Souza, A. Moura, F. R. C. Sousa and G. Fenner, "Assessing cloud computing SaaS adoption for enterprise applications using a Petri net MCDM framework," 2014 IEEE Network Operations and Management Symposium (NOMS), Krakow, 2014, pp. 1-6.

[2] M. Ruehl, T. Stefan, Malte Rupprecht, Bjorn Morr, Matthias Reinhardt, and Stephan A. W. Verclas, "Mixed-Tenancy in the Wild - Applicability of Mixed-Tenancy for Real-World Enterprise SaaS-Applications," 2014 IEEE 7th International Conference on Cloud Computing (2014): 865-72. Web.

[3] P. Jamshidi, C. Pahl, N. C. Mendonça, J. Lewis and S. Tilkov, "Microservices: The Journey So Far and Challenges Ahead," in IEEE Software, vol. 35, no. 3, pp. 24-35, May/June 2018.

[4] B. B. Rad, H. J. Bhatti & M. Ahmadi, "An introduction to docker and analysis of its performance," International Journal of Computer Science and Network Security (IJCSNS), 2017, 17(3), 228-235.

[5] Leite Leonardo, Rocha Carla, Kon Fabio, Milojicic Dejan, Meirelles Paulo, "A Survey of DevOps Concepts and Challenges," ACM Computing Surveys (CSUR), 10 December 2019, Vol.52(6), pp.1-35.

[6] Fei Hu, Meikang Qiu, Jiayin Li, "A Review on Cloud Computing: Design Challenges in Architecture and Security," December 2010, Journal of Computing and Information Technology 19(1):25-55, DOI: 10.2498/cit.1001864.

[7] Ruze Richards, Jeffrey Jackovich, "Machine Learning with AWS." Book, Packt Publishing, November 2018.

[8] Ignacio Arnaldo, Kalyan Veeramachaneni, Andrew Song, Una-May, "Bring Your Own Learner: A Cloud-Based, Data-Parallel Commons for Machine Learning," O'Reilly, IEEE Computational Intelligence Magazine 10, Feb. 2015, 20–32.

[9] Pasquale Salza Erik Hemberg, Filomena Ferrucci, Una-May O'Reilly , "cCube: A Cloud Microservices Architecture for Evolutionary Machine Learning Classification," GECCO '17: Proceedings of the Genetic and Evolutionary Computation Conference Companion, July 2017, Pages 137–138 https://doi.org/10.1145/3067695.3076089

[10] N. Huber, M. Von Quast., M. Hauck, S. Kounev "Evaluating and modeling virtualization performance overhead for cloud environments," CLOSER 2011, Proceedings of the 1st International Conference on Cloud Computing and Services Science, 2011, pp.563-573 .

[11] M. Zaharia, A. Chen, A. Davidson, A. Ghodsi, S. A. Hong, A. Konwinski, et al, "Accelerating the Machine Learning Lifecycle with MLflow" Data Engineering, vol. 39, 2018.

[12] J. Dunn. Introducing FBLearner Flow: Facebook's AI backbone. https://code.fb.com/core-data/introducing-fblearner-flow-facebook-s-ai-backbone/.

[13] J. Hermann and M. D. Balso. Meet Michelangelo: Uber's machine learning platform. https://eng.uber.com/michelangelo/.

[14] D. Baylor, E. Breck, H.-T. Cheng, N. Fiedel, C. Y. Foo, Z. Haque, S. Haykal, M. Ispir, V. Jain, L. Koc, C. Y. Koo, L. Lew, C. Mewald, A. N. Modi, N. Polyzotis, S. Ramesh, S. Roy, S. E. Whang, M. Wicke, J. Wilkiewicz, X. Zhang, and M. Zinkevich. Tfx: "A tensorflow-based production-scale machine learning platform". In Proceedings of the 23rd ACM SIGKDD International Conference on Knowledge Discovery and Data Mining, KDD '17, pages 1387–1395, New York, NY, USA, 2017. ACM.

[15] Ian Robinson, "Consumer-Driven Contracts: A Service Evolution Pattern," Jun 2006. https://martinfowler.com/articles/consumerDrivenContracts.html

[16] Bień, M., Castellotti, R., & Canali, L. "Big Data Analysis and Machine Learning at Scale with Oracle Cloud Infrastructure". 2019.

[17] Zaharia, Matei, et al. "Apache spark: a unified engine for big data processing." Communications of the ACM 59.11 (2016): 56-65.

[18] Bisong, Ekaba. "Kubeflow and Kubeflow Pipelines." Building Machine Learning and Deep Learning Models on Google Cloud Platform. Apress, Berkeley, CA, 2019. 671-685.

[19] Varanasi B., Belida S. "Documenting REST Services. In: Spring REST". Apress, 2015, Berkeley, CA

[20] George Sammons, "Introduction to AWS (Amazon Web Services) Beginner's Guide Book: Learning the basics of AWS in an easy and fast way". 2016. CreateSpace Independent Publishing Platform, North Charleston, SC, USA.

[21] Wilder, Bill. Cloud architecture patterns: using Microsoft azure. " O'Reilly Media, Inc.", 2012.

[22] Ali S. "Hosted Cloud Solutions Using Google Cloud Platform. In: Practical Linux Infrastructure". 2015. Apress, Berkeley, CA

[23] Jakóbczyk, Michał Tomasz. "Practical Oracle Cloud Infrastructure: infrastructure as a service, autonomous database, managed Kubernetes, and serverless.", 2020.

[24] Jakóbczyk, M. T. "Autonomous Database. In Practical Oracle Cloud Infrastructure" (pp. 347-408). 2020. Apress, Berkeley, CA.

[25] Shah, Jay, and Dushyant Dubaria. "Building modern clouds: using docker, kubernetes & Google cloud platform." 2019 IEEE 9th Annual Computing and Communication Workshop and Conference (CCWC). IEEE, 2019.

[26] Jae Choi, Derek L. Nazareth & Hemant K. Jain, "Implementing Service-Oriented Architecture in Organizations," Journal of Management Information Systems, 2010, 26:4, 253-286, DOI: 10.2753/MIS0742-1222260409

**Venkata Duvvuri**

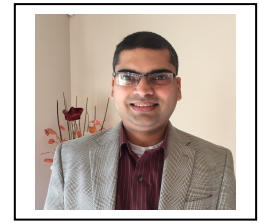

Venkata Duvvuri is a doctoral student at Department of Technology Leadership and Innovation at Purdue University. Additionally, he is an Architect level Consulting Member of Technical Staff - Data Scientist at Oracle corporation in Redwood City, CA, USA. He loves teaching and is an adjunct faculty at Northeastern University. He has held several leadership positions in data science at various companies. He holds a Master's degree in computer science from University of Massachusetts – Amherst and an MBA from University of California - Davis.